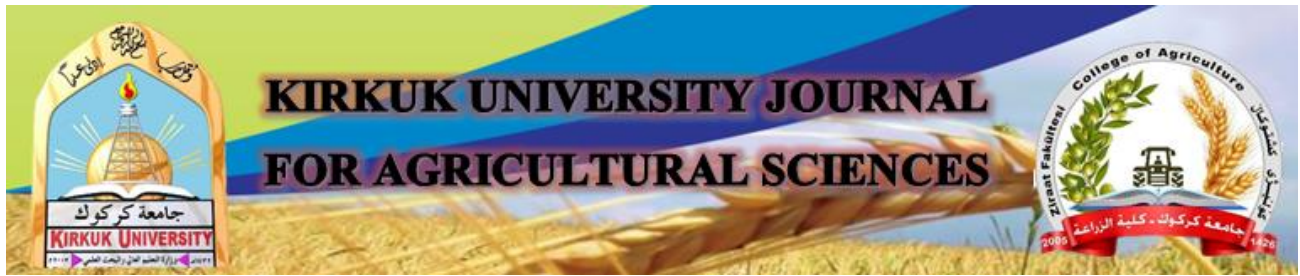

# Analysis of genetic diversity among some Iraqi durum wheat cultivars revealed by different molecular markers

Mihraban Sharif Maeruf[1]

mihraban.maeruf@univsul.edu.iq

Djshwar Dhahir Lateef[1]

djshwar.lateef@univsul.edu.iq

Kamil Mahmood Mustafa[1]

shoxan.sleman@univsul.edu.iq

Hero Fatih Hamakareem[1]

hero.hamakareem@univsul.edu.iq

Shang Hasseb Abdulqadir[1],

shang.abdalqadr@univsul.edu.iq

Dastan Ahmad Ahmad[1]

dastan.ahmad@univsul.edu.iq

Shokhan Mahmood Sleman[1]
shoxan.sleman@univsul.edu.iq

Kamaran Salh Rasul[2]
kamaran.rasul@univsul.edu.iq

[1]Biotecnology and crop science department College Of Agricultural Engineering Sciences, University of Sulaimani46001,sulaymaniyah,Kurdistan Region , Iraq.
[2]Horticulture Department, College Of Agricultural Engineering Sciences, University of Sulaimani,sulaimania 46001,Sulaymaniyah,Kurdistan Region, Iraq.

**Abstract**

Durum wheat has been cultivated since the beginning of crop domestication, occupying now the tenth ranking among the global most significant cultivated crops. Despite the fact that, the extent of the crop genetic diversity has not yet fully incorporated into modern varieties through breeding programs. In this study, a total of 35 markers (11 RAPD, 12 ISSR, and 12 CDDP) were utilized to assess the genetic variability and population structure of sixteen different cultivars of Iraqi durum wheat. Out of 294 bands obtained, 171 were identified as polymorphic: 47.00 polymorphic alleles from 98 RAPD bands, 53 polymorphic alleles from a total of 89 ISSR bands, and 71 alleles from 107 CDDP bands. The average number of observed alleles (Na), effective number of alleles (Ne), Shannon's information index (I), expected heterozygosity or gene diversity (He), unbiased expected heterozygosity (uHe), and polymorphic information content (PIC) (1.45, 1.38, 0.32, 0.22, 0.24, and 0.28, respectively) were obtained for RAPDs , (1.63, 1.45, 0.40, 0.27, 0.29, and 0.32, respectively) ISSRs and (1.35, 1.35, 0.31, 0.21, 0.23, and 0.30, respectively) for the CDDP markers. A dendrogram of two main clades (unweighted pair group method with arithmetic mean; UPGMA) and three populations of structure analysis, were obtained based on the three markers' data. The analysis of molecular variance indicated 97.00%, 97.00%, and 90.00% variability within populations, applying RAPD, ISSR, and CDDP markers, respectively. The highest diversity indices were revealed in population 2 under the RAPD and CDDP markers, whereas population 1 had the highest values of these indices according to the ISSR markers. The results provide greater knowledge on the genetic makeup of Iraqi durum wheat cultivars, that facilitate future breeding programs of this crop.

**Introduction**

Wheat, which includes common and durum varieties, is a significant grain crop widely consumed by humans globally. Durum wheat (*Triticum turgidum* L. subsp. *durum*) is a primary cereal crop that is the sole tetraploid (AABB, 2n D 4x D 28) variety of wheat used in Mediterranean regions [1]. Currently, this area contributes approximately 60% of worldwide productivity and holds the most significant genetic diversity in durum wheat germplasm [2]. Drum wheat plants exhibit a notable range of genetic and morphological characteristics, but the evolution of these plants during domestication has not been thoroughly examined or documented [3]. It is a prevalent grain crop grown by farmers, and it is grown on almost 16 million hectares globally, yielding a total of 38 million ton [4]. Wheat is widely cultivated in Iraq, and the cultivated area is estimated to be approximately 1.583 million hectares, with a total production of 4.343 million metric tons per year and an average yield of 2.744 metric tons per hectare [5]. Durum wheat is a vital food source with high levels of carbohydrates, supplementary proteins, calcium, calories, zinc, fiber, and lipids [6]. In the present day, it is used to make a range of food items including bread, cakes, pasta, biscuits, and bakery products. The rising popularity of confectionery items has led to an increased need for a diverse selection of durum wheat cultivars [7].

Breeders must enhance their production while reducing agriculture's environmental effect to deal with this challenging situation. Because the demand for wheat is expected to rise by more than 80% by 2050 compared to the current level [8]. Assessing the genetic diversity in germplasm collections is crucial for breeding programs. This is because it helps identify cultivars and lines with a wider range of features and enhanced performance under specific conditions. Successful breeding programs rely on substantial genetic variation among populations [3]. Environmental influences may influence the accuracy of morphological and biochemical indicators, but they remain valuable for assessing the genetic diversity of a population [9]. The rising reliability, consistency, and standardization of molecular markers have led to their increased utilization. Molecular markers are valuable tools for analyzing agricultural development, especially for assessing genetic diversity among crop species and their interrelationships. Molecular markers such as Random Amplified Polymorphic DNA (RAPD), Amplified Fragment Length Polymorphisms (AFLPs), Simple Sequence Repeats (SSRs), Inter Simple Sequence Repeats (ISSRs), and others are assessed using polymerase chain reaction (PCR), a commonly used technique [10]. The RAPD marker evaluation yields several markers for comparing the genotypes of distinct individuals. This method allows for large-scale screens of breeding populations and genetic resources without compromising efficiency due to its straightforward management and great economic efficiency [11]. ISSR markers can be identified by using primers that target repetitive DNA sequences and amplifying the resulting product containing SSRs. Various plant species have been used to examine genetic diversity [12]. This method is efficient for evaluating the genetic diversity of various crops because to its high repeatability, polymorphism, and informativeness [10]. In the recent years, several new alternative opportunities and powerful marker techniques have been formed. Conserved DNA-derived polymorphism (CDDP) molecular markers employ conserved DNA sequences. This approach amplifies the entire plant genome by creating specialized primers that target conserved regions of functional genes to produce informative bands. Collard and Mackill [13] describe this method as very adaptable, polymorphic, and known for its exceptional stability and reproducibility. Many researchers use this method to evaluate genetic diversity, especially those focusing on plant species [14–16]. In current study, the RAPD, ISSR, and CDDP molecular techniques were used to estimate the genetic variation of sixteen Iraqi durum wheat plants. Moreover, comprehending the genetic distances among the cultivars under study would assist in choosing genetically diverse and complementing breeding parents for the durum wheat breeding program.

**Materials and methods**
**1. Plant Materials and DNA Extraction**

Sixteen durum wheat cultivars (G1: Cham 1, G2: Ofanto, G3: Miki 3, G4: Gidra, G5: Omnabi 5, G6: Halio, G7: Waha, G8: Simeto, G9: Bohooth 5, G10: Crezo, G11: Fadda 98, G12: Acsad 65, G13: Secondroue, G14: Guayacan, G15: Zanbaka, and G16: Mosuly) were used in this study; these plants were collected from almost all agricultural research centers in Iraq, and these cultivars are the most widely cultivated in Iraq. These cultivars were grown in plastic pots (20 cm in height and 12 cm in diameter) with peatmoss, during the growing season (autumn 2022 to spring 2023) at the University of Sulaimani/College of Agricultural Engineering Sciences. Genomic DNA was isolated from the young leaves of 2-week-old plants cultivated in plastic pots using the procedure outlined by [17,18]. A NanoDrop spectrophotometer (NanoPLUS-MAANLAB AB, Sweden) and 1% agarose gel were used in order to assess both the quality and quantity of the DNA that was isolated. DNA samples were diluted to 50 ng $\mu L^{-1}$ with dd $H_2O$ and stored at -20°C until use in polymerase chain reaction (PCR).

**2. RAPD, ISSR, and CDDP assays**

For the purpose of conducting a genetic diversity analysis on the durum wheat cultivars, eleven RAPD primers [19,20], 12 ISSR primers [14,21], and 12 CDDP primers [22,23] were utilized. The polymorphism rate of these primers was taken into consideration in prior research on wheat and other crops, which led to their selection. A reaction mixture comprising 5 µL of template DNA, 8 µL of PCR master mix (Hot Start Taq Master, Addbio, Korea), 4 µL of primer (Addbio, Korea), and 5 µL of deionized water was utilized to conduct PCR amplification. Initial denaturation at 94°C for 10 minutes, followed by 36 cycles of denaturation at 94°C for 1 minute, annealing at a specific temperature (depending on primer) for 1 minute, and extension at 72°C for 2 minutes; and a final extension at 72°C for 10 minutes, were executed using a MultiGene OptiMax Thermal Cycler (Labnet Company) PCR machine. Following that, gel electrophoresis was carried out in order to separate the results of the polymerase chain reaction (PCR) on 1.5% agarose gels using a solution consisting of 1X TBE (tris-base, boric acid, and EDTA) and 0.15 µg/mL ethidium bromide. the gels were exposed to a regular voltage of 85 V for 90 minutes. In order to visualize the stained gel, an ultraviolet transilluminator (Labnet Company) was utilized in order to capture the product during the processing

Table 1. Random amplified polymorphic DNA (RAPD), inter simple sequence repeats (ISSR), conserved DNA-derived polymorphism (CDDP), sequence of primers and annealing temperatures (ATs) were used.

| RAPD primers | Primer sequence of (5′ – 3′) | AT (°C) |
| --- | --- | --- |
| GEN 2-80-8 | GGCCACAGCG | 36.00 |
| OPAD-06 | AAGTGCACGG | 36.00 |
| OPAK-15 | ACCTGCCGTT | 36.00 |
| OPW-13 | CACAGCGACA | 36.00 |
| OPAW-11 | CTGCCACGAG | 36.00 |
| OPY-04 | GGCTGCAATG | 36.00 |
| OPAR-16 | CCTTGCGCCT | 36.00 |
| OPAT-08 | TCCTCGTGGG | 36.00 |
| OPAU-12 | CCACTCGTCT | 36.00 |
| OPW-08 | GACTGCCTCT | 36.00 |
| OPP-05 | CCCCGGTAAC | 36.00 |
| ISSR Primers | | |
| UBC-808 | AGAGAG AGAGAGAGAGC | 50.00 |
| UBC-813 | CTCTCTCTCTCTCTCTT | 50.00 |
| UBC-815 | CTCTCTCTCTCTCTCTG | 50.00 |
| UBC-818 | CACACACACACACACAG | 52.80 |
| UBC-822 | TCTCTCTCTCTCTCTCA | 50.00 |
| UBC-823 | TCTCTCTCTCTCTCTCC | 50.00 |
| UBC-826 | ACACACACACACACACC | 50.00 |
| UBC-834 | AGAGAGAGAGAGAGAGGT | 50.00 |
| UBC-840 | GAG AGA GAG AGA GAG | 46.30 |
| UBC-852 | GATAGATAGACAGACA | 48.00 |
| UBC-856 | ACACACACACACACACYA | 42.00 |
| UBC-891 | ACTACTACTTGTGTGTGTGTG | 52.00 |
| CDDP Primers | | |
| WRKY-R3 | GCASGTGTGCTCGCC | |
| WRKY-R3B | CCGCTCGTGTGSACG | 50.00 |
| ABP1-1 | ACSCCSATCCACCGC | 50.00 |
| MADS-1 | ATGGGCCGSGGCAAGGTGC | 50.00 |
| WRKYR1 | GTGGTTGTGTCTTGCC | 50.00 |
| WRKYF1 | TGGCGSAAGTACGGCCAG | 50.00 |
| Myb1 | GGCAAGGGCTGCCGC | 50.00 |
| Myb2 | GGCAAGGGCTGCCGG | 50.00 |
| ERF1 | CACTACCCCGGSCTSCG | 50.00 |
| ERF2 | GCSGAGATCCGSGACCC | 50.00 |
| Knox1 | AAGGGSAAGCTSCCSAAG | 50.00 |
| Knox2 | CACTGGTGGGAGCTSCAC | 50.00 |

### 3. The process of scoring and analyzing statistical data

After amplifying the fragments, the bands were coded manually by assigning a score of 0 for no bands and a score of 1 for bands present. The Jaccard coefficient's similarity coefficient was calculated by analyzing the scored data matrices using XLSTAT 2020 computer software [24]. The Jaccard coefficient was converted into a dissimilarity matrix using the unweighted pair group technique with arithmetic averages (UPGMA) to conduct cluster analysis on cultivars. The software Power 5 Marker version 3.25 was used to calculate the polymorphism information content (PIC), allele frequency, and gene diversity. The software STRUCTURE, version 2.3.4, was used for model analysis [25]. The number of populations that were assumed to exist (K) was set anywhere from one to 10, and the method of analysis was carried out three times. Burn-in and MCMC were both set at 50,000 for each category, and the number of iterations was established at 5000. Both of these values were fixed for the purpose of establishing cultivars into populations, the run with the highest likelihood was utilized [23].

**Results**
**1. Polymorphism and distinguishing features of the RAPD, ISSR, and CDDP markers**

There was a total of 35 primers used to identify the genetic variability among the 16 Iraqi durum wheat cultivars. 11 RAPD primers, 12 ISSR primers, and 12 CDDP primers were used. The results showed that there were 294 bands, with 171 polymorphic bands. (Table 2). The RAPD primers produced 98 bands, and 47 of them were polymorphic bands, and the ISSR primers created 53 polymorphic bands out of 89 bands, while 71 polymorphic bands from 107 bands were produced by the CDDP markers. The average polymorphic bands of RAPD, ISSR, and CDDD primers were 4.27, 4.42, and 5.92, respectively. The overall number of polymorphic bands (TPB) ranged from 1 (GEN 2-80-8) to 8 (OPW-13), from 2 (UBC-813, UBC-815, UBC-822, and UBC-891) to 10 (UBC-856), and from 2 (Myb2) to 11 (ERF1) for the RAPD, ISSR, and CDDP primers, respectively.

Among the RAPD markers, the primer OPW-08 had the highest number (13.00 bands) of total bands (TB), and both primers, OPAK-15 and OPAW-11, had the lowest TB, with 6.00 bands. OPW-13 and GEN 2-80-8 presented the highest and lowest polymorphic bands (TPB) at 1.00 and 8.00, respectively. OPAU-12 revealed the maximum values of the number of observed alleles (Na), effective number of alleles (Ne), Shannon's information index (I), expected heterozygosity or gene diversity (He), and unbiased expected heterozygosity (uHe) with 1.79, 1.57, 0.48, 0.33, and 0.37, respectively,
.

while OPAW-11 observed the lowest values of Na, Ne, I, He, and uHe at 1.00, 1.06, 0.11, 0.05, and 0.05, respectively. The highest polymorphic information content (PIC) (0.48) was recorded for OPAR-16, while OPAW-11 and OPP-05 had the lowest PIC (0.17). Among the ISSR markers, the highest and lowest values of TB were indicated by UBC-852 (14 bands) and UBC-813 (2 bands), respectively. The highest PIC value was recorded for UBC-856, with 10.00 polymorphic bands, while the lowest polymorphic bands (2.00) were indicated by UBC-813, UBC-815, UBC-822, and UBC891. There was a maximum PIC value reported for UBC-856, which was 0.46, while the greatest values of Na, Ne, I, He, and uHe were recorded for UBC-822, which were 2.00, 1.78, 0.62, 0.43, and 0.47, respectively. UBC-808 revealed the lowest values of 1.29, 1.23, 0.22, 0.14, 0.15, and 0.20 for Na, Ne, I, He, uHe, and PIC, at, respectively.

Among the CDDP primers, the minimum and maximum values of TB were recorded MADS-1 (3.00 bands) and for ERF1 (15.00 bands), respectively. The maximum value of TPB was indicated by ERF1, which had 11.00 bands, and the lowest polymorphic bands (2.00 bands) were found for MADS-1 and Myb2. WRKY-R3 (1.64) and ERF2 (1.13) revealed the maximum and minimum of Na values, respectively. Among these primers, the values of Ne, I, He, uHe, and PIC that were found in Myb2 were the highest, with values of 1.48, 0.41, 0.28, 0.32, and 0.40, respectively. However, the Ne and I value revealed by ERF1 and Knox1 were 1.26 and 0.25, respectively. Additionally, ERF1 and Knox1, with 0.17 and 0.18, revealed the lowest of He and uHe values, respectively, and the lowest PIC value was indicated by ABP1-1, at 0.16.

The average values for Na, Ne, I, He, uHe, and PIC were 1.45, 1.38, 0.32, 0.22, 0.24, and 0.28 for RAPD markers, respectively, 1.63, 1.45, 0.40, 0.27, 0.29, and 0.32 for ISSR markers, respectivel, and 1.35, 1.35, 0.31, 0.21, 0.23, and 0.30 for CDDP markers, respectively

Table 2. Total number of polymorphic bands (TPB), total number of bands (TB), number of observed alleles (Na), effective number of alleles (Ne), expected heterozygosity or gene diversity (He), Shannon's information index (I), unbiased expected heterozygosity (uHe), and polymorphic information content (PIC) in wheat genotypes obtained using random amplified polymorphic DNA (RAPD), inter simple sequence repeats (ISSR), and conserved DNA-derived polymorphism (CDDP)

| RAPD markers | TB | TPB | Na | Ne | I | He | uHe | PIC |
|---|---|---|---|---|---|---|---|---|
| GEN 2-80-8 | 8.00 | 1.00 | 1.50 | 1.32 | 0.29 | 0.20 | 0.20 | 0.23 |
| OPAD-06 | 9.00 | 3.00 | 1.50 | 1.36 | 0.32 | 0.21 | 0.24 | 0.29 |
| OPAK-15 | 6.00 | 2.00 | 1.50 | 1.41 | 0.32 | 0.22 | 0.23 | 0.25 |
| OPW-13 | 10.00 | 8.00 | 1.69 | 1.47 | 0.40 | 0.27 | 0.31 | 0.29 |
| OPAW-11 | 6.00 | 2.00 | 1.00 | 1.06 | 0.11 | 0.05 | 0.05 | 0.17 |
| OPY-04 | 10.00 | 5.00 | 1.60 | 1.45 | 0.37 | 0.25 | 0.28 | 0.32 |
| OPAR-16 | 10.00 | 3.00 | 1.00 | 1.35 | 0.30 | 0.20 | 0.21 | 0.48 |
| OPAT-08 | 9.00 | 7.00 | 1.39 | 1.35 | 0.28 | 0.20 | 0.21 | 0.24 |
| OPAU-12 | 10.00 | 7.00 | 1.79 | 1.57 | 0.48 | 0.33 | 0.37 | 0.23 |
| OPW-08 | 13.00 | 7.00 | 1.50 | 1.50 | 0.40 | 0.28 | 0.30 | 0.38 |
| OPP-05 | 7.00 | 2.00 | 1.50 | 1.38 | 0.31 | 0.22 | 0.22 | 0.17 |
| Mean | 8.91 | 4.27 | 1.45 | 1.38 | 0.32 | 0.22 | 0.24 | 0.28 |
| Total | 98.00 | 47.00 | | | | | | |
| ISSR markers | | | | | | | | |
| UBC 808 | 9.00 | 7.00 | 1.29 | 1.23 | 0.22 | 0.14 | 0.15 | 0.20 |
| UBC 813 | 2.00 | 2.00 | 1.50 | 1.41 | 0.32 | 0.22 | 0.23 | 0.28 |
| UBC 815 | 6.00 | 2.00 | 1.50 | 1.33 | 0.33 | 0.21 | 0.23 | 0.36 |
| UBC 818 | 6.00 | 5.00 | 1.80 | 1.42 | 0.43 | 0.27 | 0.30 | 0.29 |
| UBC 822 | 5.00 | 2.00 | 2.00 | 1.78 | 0.62 | 0.43 | 0.47 | 0.42 |
| UBC 823 | 9.00 | 3.00 | 1.50 | 1.47 | 0.39 | 0.26 | 0.28 | 0.39 |
| UBC 826 | 6.00 | 3.00 | 1.67 | 1.41 | 0.39 | 0.25 | 0.28 | 0.33 |
| UBC 834 | 8.00 | 4.00 | 1.33 | 1.28 | 0.28 | 0.18 | 0.19 | 0.28 |
| UBC 840 | 8.00 | 4.00 | 1.50 | 1.48 | 0.38 | 0.27 | 0.29 | 0.21 |
| UBC 852 | 14.00 | 9.00 | 1.78 | 1.58 | 0.48 | 0.33 | 0.35 | 0.35 |
| UBC 856 | 12.00 | 10.00 | 1.90 | 1.64 | 0.53 | 0.37 | 0.40 | 0.46 |
| UBC 891 | 4.00 | 2.00 | 1.75 | 1.43 | 0.40 | 0.26 | 0.29 | 0.27 |
| Mean | 7.42 | 4.42 | 1.63 | 1.45 | 0.40 | 0.27 | 0.29 | 0.32 |
| Total | 89.00 | 53.00 | | | | | | |
| CDDP markers | | | | | | | | |
| WRKY-R3 | 9.00 | 7.00 | 1.64 | 1.45 | 0.37 | 0.26 | 0.32 | 0.20 |
| WRKY-R3B | 10.00 | 5.00 | 1.50 | 1.43 | 0.36 | 0.25 | 0.28 | 0.34 |
| ABP1-1 | 6.00 | 6.00 | 1.33 | 1.29 | 0.26 | 0.18 | 0.18 | 0.16 |
| MADS-1 | 3.00 | 2.00 | 1.25 | 1.34 | 0.29 | 0.20 | 0.20 | 0.32 |
| WRKYR1 | 13.00 | 10.00 | 1.35 | 1.35 | 0.32 | 0.21 | 0.23 | 0.34 |
| WRKYF1 | 9.00 | 7.00 | 1.43 | 1.37 | 0.30 | 0.21 | 0.22 | 0.30 |
| Myb1 | 8.00 | 3.00 | 1.25 | 1.32 | 0.28 | 0.19 | 0.20 | 0.32 |
| Myb2 | 7.00 | 2.00 | 1.50 | 1.48 | 0.41 | 0.28 | 0.32 | 0.40 |
| ERF1 | 15.00 | 11.00 | 1.15 | 1.26 | 0.26 | 0.17 | 0.18 | 0.34 |
| ERF2 | 6.00 | 4.00 | 1.13 | 1.30 | 0.27 | 0.18 | 0.20 | 0.33 |
| Knox1 | 11.00 | 8.00 | 1.31 | 1.28 | 0.25 | 0.17 | 0.18 | 0.27 |
| Knox2 | 10.00 | 6.00 | 1.42 | 1.37 | 0.35 | 0.23 | 0.26 | 0.32 |
| Mean | 8.92 | 5.92 | 1.35 | 1.35 | 0.31 | 0.21 | 0.23 | 0.30 |
| Total | 107.00 | 71.00 | | | | | | |
| Total of all | 294 | 171 | | | | | | |

markers

## 2. Cluster analysis using RAPD, ISSR, and CDDP data

A dendrogram produced by UPGMA showed that the RAPD markers could be used to classify 16 Iraqi durum wheat cultivars into two major clades (Figure 1A). Clade 1 (in red) included 14 cultivars, and four subclades were generated; the first subclade included eight cultivars (G1, G2, G3, G4, G5, G9, G11, and G12), with only 1 cultivar composing the second subclade (G7); the third subclade contained three cultivars (G8, G13, and G16); and two cultivars (G10 and G14) generated the fourth subclade. Clade 2 (in green) included only two cultivars (G6 and G15). In accordance with the ISSR data, the UPGMA method was utilized for the purpose of conducting clade analysis on all of the cultivars, and the results showed that two primary clades were produced (Figure 1B). The first clade (red) included four cultivars (G1, G2, G15, and G16), and two subclades were generated. Whereas the second clade 2 (green) included 12 cultivars, two subclades were generated. The first subclade included 11 cultivars, while the second subclade contained only one cultivar (G8). Two main clades were generated based on the CDDP marker data (Figure 1C). The first clade (red) included 15 cultivars, and two subclades were produced; the first subclade comprised 14 cultivars, and the second subclade included only one cultivar. However, only one cultivar (G9) formed the second clade.

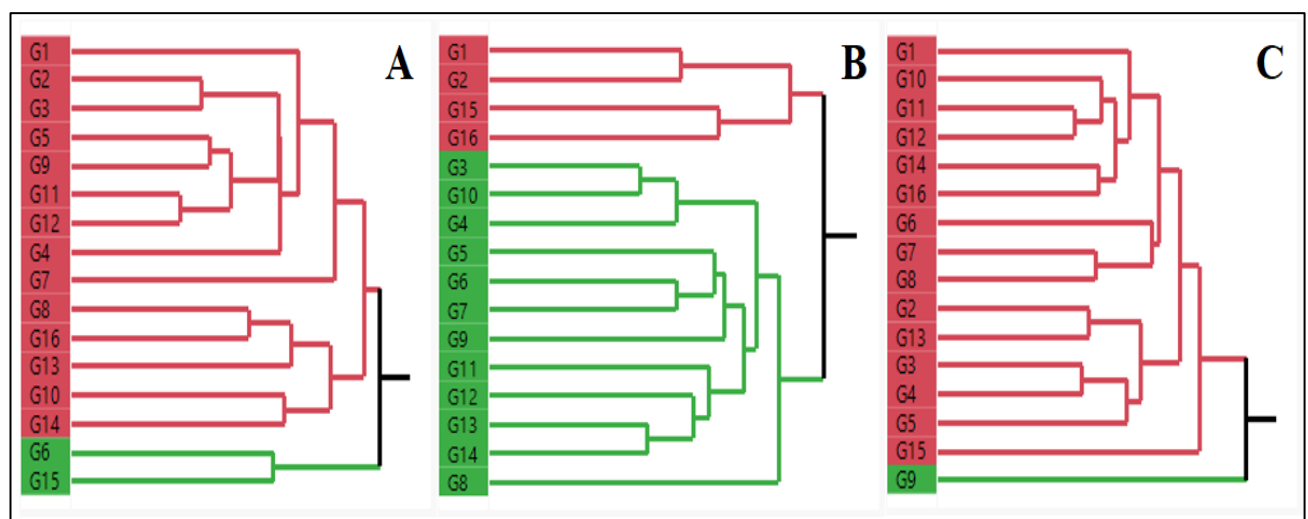

Figure 1 Unweighted pair group method with arithmetic mean (UPGMA) dendrograms of sixteen durum wheat genotypes (G1–G16) based on random amplified polymorphic DNA (RAPD), inter simple sequence repeats (ISSR), and conserved DNA-derived polymorphism (CDDP). The colors of the branches correlate to the groups that are clustering; red represents the first cluster, and green represents the second cluster.

## 3. Analyzing population structure using RAPD, ISSR, and CDDP datasets.

RAPD markers were used in structural analysis to determine how the populations of 16 types of Iraqi durum wheat were divided. Using genotypic information from the entire genome, the K value was used to calculate the number of accession clusters that were associated with the genome. The best K value was found by graphing the number of clusters (K) against K, with the largest peak appearing at K = 3. This allowed for the determination of the optimal K value (Figure 2A). It was discovered through the use of the optimal K value that three populations, namely population 1 in red, population 2 in green, and population 3 in blue, have the highest likelihood of clustering (Figure 2B). The membership probability was set at 0.80. Independently, the cultivars that comprised each subgroup were determined, and fractions that were lower than 0.20 indicated the presence of mixing. The first population included three durum wheat cultivars (G13, G14, and G16), whereas the second population included six durum wheat cultivars: G3, G4, G5, G7, G9, and G11. The third population included four cultivars, G1, G6, G10, and G15. Among the three populations, there were three cultivars that were found to be admixed, which is evidence that these cultivars are not pure.

A structure-harvester analysis of the ISSR data was used to determine the K value that was the most accurate. In the graph that was created by plotting the number of clusters (K) against K, the maximum point was found at K = 3 (Figure 2C). Three genetic groups emerged from the 16 Iraqi durum wheat cultivars (Figure 2D). The first population (red) included only two cultivars (G15 and G16); six cultivars (G3, G4, G7, G8, G10, and G11) generated the second population (green); and the third population (blue) comprised seven cultivars; G12 was considered an admixture cultivar between populations. Using STRUCTURE software, CDDP data were used to conduct population structure analysis of 16 durum wheat cultivars via an admixture model. The Evano method identified K=3 as the optimal number of clusters (Figure 2E). A cultivar was considered a pure member of a cluster if its likelihood of membership in that cluster exceeded 80%. The first population (red) included only two cultivars (G7 and G9), and six cultivars (G1, G11, G12, G14, G15, and G16) generated the second population (green), whereas the third population comprised five cultivars (G2, G3, G4, G5, and G13). There were thirteen cultivars that were deemed to be pure, whereas the remaining three cultivars (G6, G8, and G10) were categorized as admixed (Figure 2F).

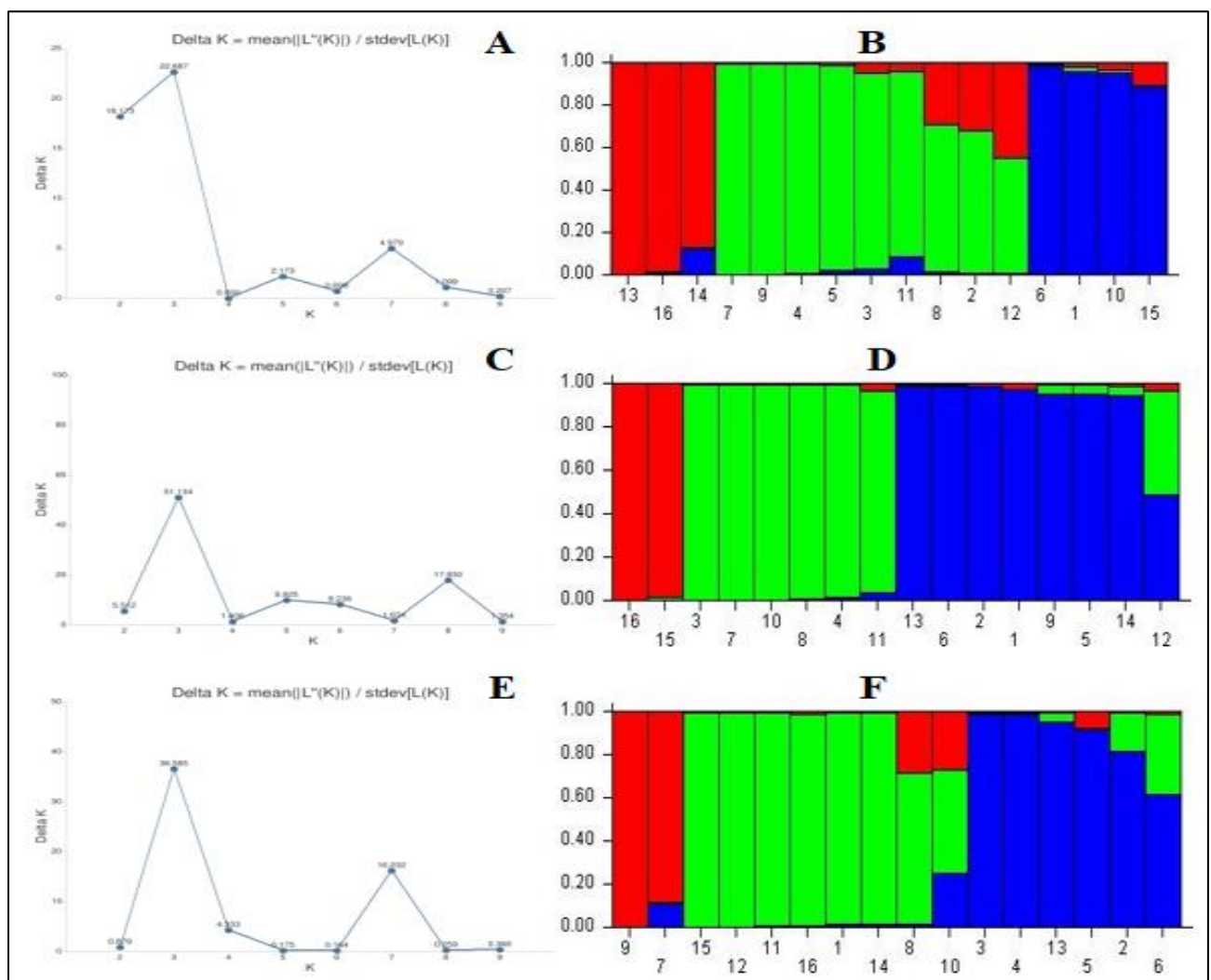

Figure 2 A delta K derived using RAPD data for a number of different population numbers (K); B evaluation of the population structure of sixteen durum wheat plants (G1–G16) based on K=3 that was determined from RAPD data; C delta K for a variety of population numbers (K) generated from information obtained from ISSR; D According to the information provided by ISSR, the population structure of sixteen durum wheat plants (G1–G16) on K=3 is anticipate.; E delta K for various population numbers (K) resulting from CDDP data; F anticipated population structure of sixteen durum wheat plants (G1–G16) on K=3 based on CDDP data

**4. Analysis of molecular variance and population diversity indices**

The RAPD, ISSR, and CDDP marker data were subjected to analysis of molecular variance (AMOVA) to determine the genetic differences between and within populations. According to RAPD, ISSR, and CDDP markers, the PhiPT values were 0.28, with a p value of 0.001; 0.24, with a p value of 0.001; and 0.06, with a p value of 0.001, respectively. The RAPD, ISSR, and CDDP marker data indicated that the overall variation among the populations described 3.00, 3.00, and 10.00%, respectively. However, the results of the RAPD, ISSR, and CDDP markers showed the greatest amount of variance within the population, with 97.00, 97.00, and 90.00%, respectively (Table 3).

Table 3 Molecular variance analysis of durum wheat cultivar populations using random amplified polymorphic DNA (RAPD), inter simple sequence repeat (ISSR), and conserved DNA-derived polymorphism (CDDP) data.

| Source | Degree of freedom | Sum of squares | Mean of squares | Estimated variance | Variation% |
|---|---|---|---|---|---|
| RAPD markers | | | | | |
| Among Pops | 1.00 | 8.27 | 8.27 | 0.26 | 3.00 |
| Within Pops | 14.00 | 102.86 | 7.35 | 7.35 | 97.00 |
| Total | 15.00 | 111.13 | | 7.61 | 100.00 |
| PhiPT | 0.28** | | | | |
| *p* Value | 0.001 | | | | |
| ISSR markers | | | | | |
| Among Pops | 1.00 | 10.90 | 10.90 | 0.28 | 3.00 |
| Within Pops | 14.00 | 129.42 | 9.24 | 9.24 | 97.00 |
| Total | 15.00 | 140.31 | | 9.52 | 100.00 |
| PhiPT | 0.24** | | | | |
| *p* Value | 0.001 | | | | |
| CDDP markers | | | | | |
| Among Pops | 1.00 | 16.38 | 16.38 | 1.36 | 10.00 |
| Within Pops | 14.00 | 162.93 | 11.64 | 11.64 | 90.00 |
| Total | 15.00 | 179.31 | | 12.99 | 100.00 |
| PhiPT | 0.06** | | | | |
| *p* Value | 0.001 | | | | |

PhiPT the fraction of the overall genetic variation that exists between the individuals within a population, p value probability value, **statistically significant at the 0.01 level
.

The diversity indices (Na, Ne, I, He, uHe, and Nm (gene flow between populations)) of the populations were revealed in Table 5. The highest diversity indices of Na, Ne, I, He, and uHe were recorded for population 1 (Pop1), with 1.96, 1.65, 0.53, 0.36, and 0.38 for the RAPD markers and 1.89, 1.54, 0.48, 0.32, and 0.33 for the CDDP markers, respectively. In contrast, population 2 (Pop2) had the highest values of these diversity indices according to the ISSR markers, with values of 1.94, 1.53, 0.48, 0.32, and 0.33, respectively. The gene flow between the populations (Nm) according to the RAPD, ISSR, and CDDP markers was 0.28, 0.24, and 0.22, respectively.

Table 4 Diversity indices of populations determined using random amplified polymorphic DNA (RAPD), inter simple sequence repeat (ISSR), and conserved DNA-derived polymorphism (CDDP) data.

| Pop | Na | Ne | I | He | uHe | Nm |
|---|---|---|---|---|---|---|
| RAPD markers | | | | | | |
| Pop1 | 1.96 | 1.65 | 0.53 | 0.36 | 0.38 | 0.28 |
| Pop2 | 1.06 | 1.20 | 0.17 | 0.12 | 0.16 | |
| ISSR markers | | | | | | |
| Pop1 | 1.38 | 1.41 | 0.34 | 0.23 | 0.27 | 0.24 |
| Pop2 | 1.94 | 1.53 | 0.48 | 0.32 | 0.33 | |
| CDDP markers | | | | | | |
| Pop1 | 1.89 | 1.54 | 0.48 | 0.32 | 0.33 | 0.22 |
| Pop2 | 0.82 | 1.16 | 0.14 | 0.10 | 0.13 | |

*Pop* population, *Na is the* number of observed alleles, *Ne* is the effective number of alleles, *I is* Shannon's information index, *He* is the expected heterozygosity or gene diversity, *uHe* is the unbiased expected heterozygosity, and *Nm is* the number of migrants per generation or gene flow between populations.

**Discussion**

In a period of climate change and evolving end-user demands, genetic diversity is crucial for providing new traits and alleles for plant breeding, especially to meet the uncertain challenges that lie ahead. Instead, to more effectively focus on their crossing schemes, breeders would benefit from knowing which optimal sources of variety should be incorporated into each program [26]. In addition, the evaluation of genetic variety and the analysis of population structure both have a multitude of advantages for the preservation of genetic resources and the implementation of an efficient breeding program [14]. A more precise evaluation of the genetic diversity present in a particular population group is associated with an increase in the capability of recognizing a superior cultivar.

In the current study, eleven RAPD, twelve ISSR, and twelve CDDP markers distributed across all the wheat chromosomes were used to group sixteen Iraqi durum wheat cultivars. The use of many molecular markers that are dispersed throughout the entire genome of a crop has the advantage of ensuring that all sections of the genome are given equal chances of representation. This, in turn, prevents erroneous assessments of the genetic similarities and differences that exist between individuals [27]. An examination into genetic diversity and population genetics was carried out with the help of the genotypic data of the cultivars that were collected. This investigation could potentially serve as the foundation for future breeding efforts in Iraqi durum wheat. A total of 35 diverse primers amplified 171 polymorphic products, and the average number of polymorphic bands produced by CDDP primers from various markers was 5.92, which was higher than the value produced by RAPD primers (4.27), as well as the value produced by ISSR primers (4.41). Previous research conducted on the genetic diversity of Iraqi durum wheat using RAPD markers revealed a lower average number of polymorphic bands (2.5). These variations most likely result from polymorphisms in the various genomic regions of durum wheat [28]. A similar result was detected by using fifteen ISSR primers, which amplified a total of 221 bands in a set of 41 durum wheat accessions, 163 of which were polymorphic[29]. As opposed to ISSR markers, which target regions flanking microsatellites in the genome, CDDP tags amplify sections of functional genes that are conserved [30]. The high degrees of polymorphism displayed by the different markers made it evident that they could be applied to studies of genetic variability in durum wheat cultivars [31].

Measures of genetic variety among genotypes in breeding populations, known as the gene diversity indices (Na, Ne, I, He, uHe, and PIC), provide insight into the evolutionary pressures acting on alleles and the rate of mutation a locus may have experienced throughout time [3]. PIC values are reliable indicators of marker significance for relationship analysis when evaluating the degree of inheritance between offspring and their parents' genotypes. [32]. Several other indices provide estimates of average heterozygosity and genetic distance among individuals in a group, as well as information on gene diversity of haploid markers [33].

The average PIC values for the RAPD, ISSR, and CDDP markers in our study were 0.28, 0.32, and 0.30, respectively. The values were established by comparison. The Iraqi durum wheat cultivars showed significant genetic diversity, as indicated by the results of three markers. The markers' values exceeded 0.25, signifying considerable genetic variation. Markers with a PIC value below 0.25 were deemed somewhat useful, but markers with a PIC value above 0.5 were considered highly informative. PIC values ranging from 0.25 to 0.5 were deemed moderately beneficial [34–36]. A similar set of data was published by Eltaher et al. [37] with regard to winter wheat. In terms of their ability to differentiate between individuals, the three different types of markers have the following order of discrimination power: CDDP > ISSR > RAPD.

Diversity indices in statistics are utilized to describe the variability present in a population where each individual is part of a distinct group. The indices that have higher values indicate a greater diversity, whilst the indices that have lower values imply a lower diversity. The expected heterozygosity (He) values for RAPD, ISSR, and CDDP had mean values of 0.22, 0.27, and 0.21, respectively. These values indicate a modest level of genetic variability within a population among its cultivars. The value of Shannon's information index (I) that was acquired by the CDDP markers was higher than the value that was obtained by the RAPD and ISSR markers on average. Based on these data, it was determined that the CDDP genome has a significant amount of diversity between the cultivars. The framework for comprehending population structure has been developed by examining the diversity of durum wheat cultivars. Analysis of population structure helps to explain genetic diversity and facilitates association mapping research in the future. A mapping population with population structure may produce erroneous positive correlations between features and markers [37]. The genetic variety and accuracy of genome-wide association studies (GWASs) can be improved by the use of population-distinguishing analysis.

In our investigation, two major clusters were produced based on all the markers; the distance between the cultivars in the cluster may be useful for breeding programs because more cultivars are separated; for example, the Cham 1 cultivar has the greatest genetic distance from the Simeto, Bohooth 5, and Guayacan cultivars. Similar results were revealed in durum wheat by Alemu et al. [38]. Another study using ISSR markers for clustering 41 durum wheat accessions and four groups was produced [29]. Structural analysis revealed that, unlike in the dendrogram plot, the populations may be categorized into three distinct groups. In the dendrogram plot, the cluster distribution took on a pattern that was strikingly similar to the distribution pattern that was found in the structure analysis. Due to the presence of admixed accessions, it was determined that the majority of the diversity was located inside the clusters. Historically, it was believed that the process of ancestral mixing was connected to the exchange of plant germplasm and hybridization. Alemu et al. [38] reported similar results in which three populations were generated among 167 durum wheat cultivars by using SNP markers. The study showed differences among and between the populations tested using the AMOVA statistic. Interestingly, in this study, the intrapopulation variation exceeded the interpopulation variation. Gene drift and gene flow significantly affect the studied populations of these durum wheat cultivars. A previous study suggested that the substantial variance within populations can be attributed to natural crossing, clonal proliferation, or long-lived cultivars [39]. Diversity indices diversity between the populations. Two populations were created, and the diversity indices were greater in population 2 for both the RAPD and CDDP markers. Additionally, the Nm values were 0.28, 0.24, and 0.22 for the RAPD, ISSR, and CDDP markers,

**Conclusion**

Sixteen durum wheat samples were collected and genotyped using 35 RAPD, ISSR, and CDDP markers to assess genetic diversity and population structure within the cultivars. The data obtained from marker analysis could be more effective at determining genetic variation among durum wheat varieties. Durum wheat cultivars displayed significant heterogeneity in the genetic markers RAPD, ISSR, and CDDP, which could be valuable for genetic research and breeding initiatives. Diversity indices revealed the presence of moderate genetic diversity in Iraqi durum wheat cultivars. The obtained average PIC values ordered the power of discrimination of durum wheat as follows: ISSR>CDDP>RAPD. Dendrograms were used to divide the cultivars into two clades, and three populations were created for structural analysis based on all the markers. These analyses can assist in assessing the variety of germplasm collections and choosing cultivars for future breeding. Hence, sustainable conservation and utilization of durum wheat genetic resources are key for future breeding strategies in Iraq.

**Conflicts of interest:** The authors declare that they have no known competing financial interests or personal relationships that could have appeared to influence the work reported in this paper.

**Acknowledgments:** The authors are grateful to the people who worked at the College of Agricultural Engineering Sciences at the University of Sulaimani for their assistance and cooperation throughout the research process.

**Funding:** This research received no external funding.

**Ethical approval:** This study is not applicable

**Consent to participate:** Not applicable among populations revealed the genetic respectively. A Nm value less than one indicates limited gene exchange among populations [37]. However, in our study, the Nm value was less than one, suggesting low genetic exchange or low gene flow among the populations

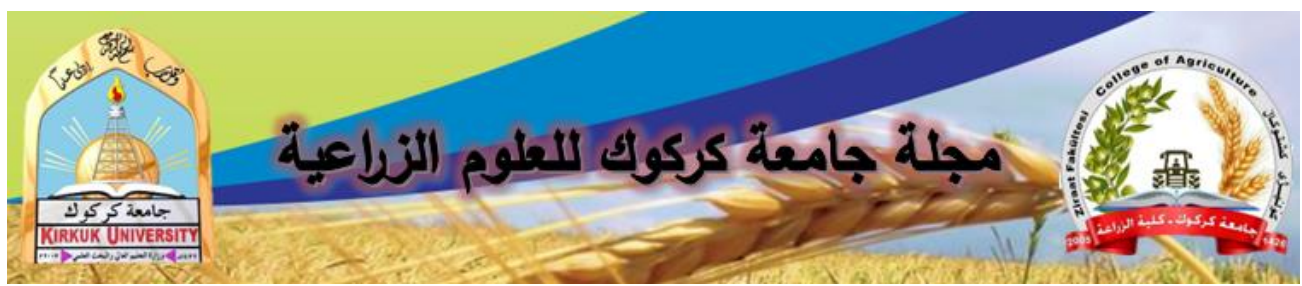

# تحليل التنوع الجيني بين بعض أصناف الحنطة الخشنة العراقية مستخدما البادئات الجزيئية المتنوعة

| مهربان شريف معروف[1] | دژوار ظاهر لطيف[1] | كامل محمود مصطفى[1] |
|---|---|---|
| mihraban.maeruf@univsul.edu.iq | djshwar.lateef@univsul.edu.iq | shoxan.sleman@univsul.edu.iq |

| هيرو فاتح حمه كريم[1] | شەنگ حسيب عبدالقادر[1] | داستان احمد احمد[1] |
|---|---|---|
| hero.hamakareem@univsul.edu.iq | shang.abdalqadr@univsul.edu.iq | dastan.ahmad@univsul.edu.iq |

| كامران صالح رسول[2] | شوخان محمود سليمان[1] |
|---|---|
| kamaran.rasul@univsul.edu.iq | shoxan.sleman@univsul.edu.iq |

[1]قسم التكنولوجيا الحياتية وعلم المحاصيل الحقلية, كلية علوم الهندسة الزراعية، جامعة السليمانية، السليمانية, العراق.
[2]قسم البستنة, كلية علوم الهندسة الزراعية، جامعة السليمانية، السليمانية, العراق.
● تاريخ استلام البحث ٢٠٢٤/٥/٣ وتاريخ قبوله ٢٠٢٤/٦/٨.

## الخلاصة

تمت زراعة الحنطة الخشنة منذ بداية استئناث المحاصيل، ويحتل الآن المرتبة العاشرة بين أهم المحاصيل المزروعة على مستوى العالم. على الرغم من أن مدى التنوع الوراثي للمحاصيل لم يتم استخدامه بشكل كامل في الأصناف الحديثة من خلال برامج التربية. في هذه الدراسة، تم استخدام مجموعه ٣٥ باديء (١١ RAPD، ١٢ ISSR، و ١٢ CDDP) لتقييم التباين الوراثي والتركيبي المجتمعي لستة عشر صنفًا مختلفًا من الحنطة الخشنة العراقية. تم الحصول على ٢٩٤ اليل ، وتم تحديد ١٧١ منهم بوليمورفك: ٤٧ أليل من RAPD ٩٨ ، و٥٣ أليل من إجمالي ISSR ٨٩، و٧١ أليل من ١٠٧ CDDP. متوسط عدد الأليلات الكلية (Na)، والعدد للأليلات الفعالة (Ne)، ومؤشر شانون (I)، والخلط التركيبى المتوقع أو التنوع الجيني (He)، والخلط التركيبى المتوقع غير متوقع (uHe)، ومحتوى المعلومات متعدد الأشكال (PIC) (١.٤٥) ، تم الحصول على ١.٣٨ ، ٠.٣٢ ، ٠.٢٢ ، ٠.٢٤ ، و ٠.٢٨ ، على التوالي) لـ RAPDS ، 1.63، ١.٤٥، ٠.٤٠ ، ٠.٢٧ ، ٠.٢٩ ، و ٠.٣٢ ، على التوالي) ISSR و (١.٣٥ ، ١.٣٥ ، ٠.٣١ ، ٠.٢١ ، ٠.٢٣ ، ٠.٣٠ ، ٠.٣٠ ، على التوالي) لباديء CDDP. تم الحصول على التحليل الهرمي مكونة من مجموعتين رئيسيتين (UPGMA) وثلاث مجموعات من تحليل الهرمي، بناءً على بيانات للبادىء الثلاث. أشار تحليل التباين الجزيئي إلى بنسبة تباين ٩٧% و٩٧% و٩٠% ضمن المجموعات العشيرية، مع تطبيق كل من RAPD وISSR وCDDP، على التوالي. تم الكشف عن أعلى مؤشرات التنوع في العشيرة ٢ مستخدما RAPD و CDDP، في حين كان لدى عشيرة ١ أعلى قيم هذه اللبادئات كانت لـ ISSR. توفر النتائج معرفة أكبر حول التركيب الوراثي لأصناف الحنطة الخشنة العراقية، مما يسهل برامج التربية المستقبلية لهذا المحصول.

**الكلمات المفتاحية:** حنطة الخشنة، تباين الوراثي، التحليل الهرمي، التكيب العشيري، التحسين الوراثي.